\documentclass[
reprint,
 amsmath,amssymb,
 aps,
]{revtex4-1}

\usepackage{graphicx}% Include figure files
\usepackage{dcolumn}% Align table columns on decimal point
\usepackage{bm}% bold math
\usepackage{amsmath}% bold math

\usepackage{color}
\begin{document}

\preprint{APS/123-QED}

\title{Five-Partite Entanglement Generation in A High-Q Microresonator}% Force line breaks with \\

\author{Yutian Wen$^1$}
\author{Xufei Wu$^1$}
\author{Rongyu Li$^1$}
\author{Qiang Lin$^2$}
\author{Guangqiang He$^1$}% \corref{cor1}}
\altaffiliation{Author to whom correspondence should be addressed. Electronic mail:gqhe@sjtu.edu.cn}
\affiliation{$^1$State Key Lab of Advanced Optical Communication Systems and Networks, Department of Electronic Engineering, Shanghai Jiao Tong University, Shanghai 200240, China\\
$^2$Department of Elctrical and Computer Engineering, University of Rochester, NY 14627, USA}
\date{\today}% It is always \today, today,
\begin{abstract}
We propose to produce five-partite entanglement via cascaded four-wave mixing in a high-Q microresonator that may become a key to future one-way quantum computation on chip. A theoretical model is presented for the underlying continuous-variable entanglement among the generated comb modes that is expansible to more complicated scenarios. We analyze the entanglement condition when the van Loock and Furusawa criteria are violated, and discuss the device parameters for potential experimental realization that may be utilized to build an integrated compact five-partite entanglement generator. The proposed approach exhibits great potential for future large-scale integrated full optical quantum computation on chip.
\begin{description}
\item[PACS numbers]
03.67.Lx, 42.65.Wi, 42.65.Yj, 03.65.Ud.
\end{description}
\end{abstract}
\maketitle

%\tableofcontents

\section{Introduction}
Quantum computation (QC) is expected to provide exponential speedup for particular mathematical problems such as integer factoring~\cite{shor1997polynomial} and quantum system simulation~\cite{feynman1982simulating}. However, any practical QC system must overcome the inevitable decoherence problem and achieve scalability.
The traditional ``circuit'' QC model keeps quantum information in a physical system where quantum memory units undergo precise controlled unitary evolution simultaneously, leading to serious scalability issue. To circumvent this challenge, an ``one-way'' quantum computation model was proposed~\cite{raussendorf2001one}, where quantum information exists virtually in a cluster state~\cite{briegel2001persistent} and one can perform any desired quantum algorithm by conducting a sequence of local measurements. With this approach, the most challenging part is now transferred from conducting the unitary operation in a large scale into the generation of a cluster state, or more generally, a universal multipartite entangled state. The aim of this paper is to investigate the possibility of a novel integrated approach for generating multipartite entangled states.

\begin{figure}
\includegraphics[width=8cm]{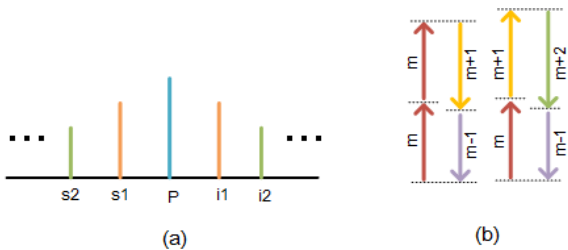}% Here is how to import EPS art
\caption{\label{fig:frequency_conversion} (a) The spectrum of an ideal frequency comb is discrete, equally spaced, and covers a wide band. (b) Energy level diagram of degenerate (left) and non-degenerate (right) FWM. }
\end{figure}

Optical frequency combs (OFCs) have been shown to be capable of preserving cluster states~\cite{menicucci2008one, roslund2013wavelength}. An OFC is a light source composed of equally spaced discrete frequency components, as illustrated in Fig.~\ref{fig:frequency_conversion} (a). Actual OFCs might extend to an extremely broad band with hundreds of frequency components~\cite{Cundiff03, Diddams10}, each of which corresponds to a comb mode (marked by a mode number, say $m$). OFCs are favorable for QC for their robustness to decoherence~\cite{roslund2013wavelength}, since photons are less likely to interact with the environment compared with other physical systems such as atoms~\cite{menicucci2008one}.

OFCs have already been utilized in many applications such as frequency metrology, telecommunications, optical and microwave waveform synthesis, and molecular spectroscopy~\cite{Cundiff03, Diddams10}. Conventionally, OFCs are generated in mode-locked lasers that are usually bulky, difficult to operate, and susceptible to environmental perturbations ~\cite{roslund2013wavelength}. It is recently reported that OFCs can also be generated from monolithic microresonators ~\cite{del2007optical, kippenberg2011microresonator} through cascaded four-wave mixing (FWM).

In a high-Q microresonator with appropriate dispersion, an intense pump wave launched into a cavity mode would excite four-wave mixing processes among different cavity modes via the optical Kerr effect \cite{kippenberg2011microresonator}. There are dominantly two types of FWM, degenerate and non-degenerate, which are illustrated in Fig.~\ref{fig:frequency_conversion} (b). Due to the momentum conservation among the interacting photons, a degenerate process converts two identical photons in a same mode at $m$ into two dissimilar photons at modes $m-1$ and $m+1$, respectively. Similarly, a non-degenerate process converts two photons from modes $m$ and $m+1$ into two new photons at modes $m-1$ and $m+2$. The iteration of these two processes thus produce an optical frequency comb~\cite{kippenberg2011microresonator}, with a spectral extent determined by the group-velocity dispersion of the device.

The beauty of such scheme lies in the nature of high-Q microresonators. First, the optical field is strongly confined inside a small volume, leading to significantly enhanced nonlinear optical interactions. Second, due to the exceptionally high quality factors ($Q$) of microresonators, the photon life time inside the cavity is much longer than that in those traditional cavities so that different frequency components have enough time to entangle with each other. Finally, the integrated chip-scale platform of microresonators exhibit great potential for eventually realizing a large-scale integrated full optical quantum computer~\cite{chembo2010modal}.

These facts inspire us to explore the potential of OFCs for producing multi-partite entangled states inside a microresonator. Although two-mode quantum squeezing has been intensively investigated for parametric processes \cite{walls2007quantum, Shelby86, Kimble86, Kimble87, Carmichael88, Mlynek95, Mlynek97, Laurat05, Vahlbruch08, Furusawa10}, the quantum properties of microresonator-based frequency comb generation has not yet been fully addressed. On the other hand, there have been both theoretical analysis~\cite{lin2006silicon, lin2007nonlinear, osgood2009engineering, chen2011frequency} and experimental investigation~\cite{sharping2006generation, takesue2008entanglement, takesue2008generation, harada2008generation, harada2010frequency} on photon pair generation inside micro/nanophotonic devices. Yet all of them focused on the bipartite discrete-variable entanglement and the bipartite methodology cannot be applied to the analysis of entanglement among three or more frequency components. In this paper, we present a theoretical model to describe the five-partite continuous-variable (CV) entanglement among frequency comb modes. We solve the Fokker-Planck equation in P representation and then analyze the entanglement condition when van Loock and Furusawa criteria are violated.
We also present an experiment scheme to verify our prediction.

The rest of this paper is arranged as follows: in section~\ref{sec: System Model}, we propose an experiment scheme to measure the continuous-variable five-partite entanglement generated from a microresonator. From section~\ref{sec: Equations Of Motion For The Full Hamiltonian} to section~\ref{sec: Five-partite Entanglement Measures}, we analyze how the degree of five-partite entanglement may vary with respect to different cavity configuration. Section~\ref{sec: Output Fluctuation Spectra} is for some discussion on the simulated output fluctuation spectra. Finally the conclusions are drawn.

\section{System Model   \label{sec: System Model}}

\begin{figure}
\includegraphics[width=8cm]{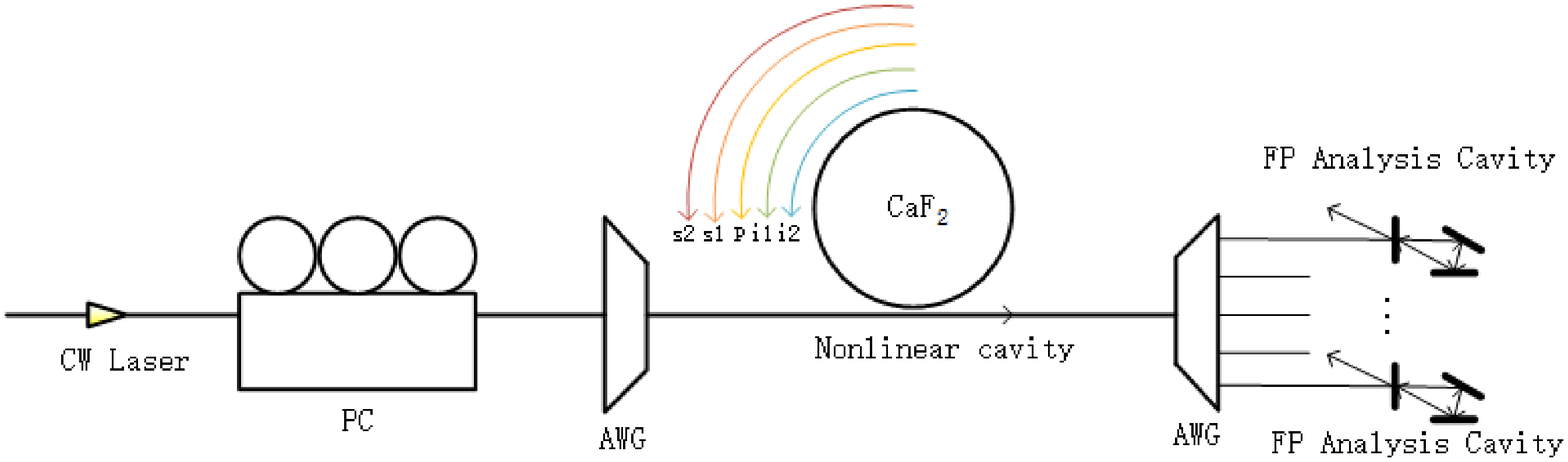}% Here is how to import EPS art
\caption{\label{fig:scheme}OFC generator with a Calcium Fluoride cavity and angle-polished fiber couplings. CW, continuous-wave; PC, polarization controller; AWG, arrayed waveguide grating.}
\end{figure}

The experiment scheme is shown in Fig.~\ref{fig:scheme}. A narrow linewidth continuous-wave laser is used to pump the system. The pump beam first go through a polarization controller, then passes an arrayed waveguide grating (AWG) which serves as a narrow bandwidth bandpass filter. The pump beam is coupled into the nonlinear dielectric cavity using evanescent fields.
The intracavity photons interact through FWM and generate the OFC, which is later extracted through evanescent fields again. Once the different frequency components of OFC is separated by AWG, we will be able to analyze each component in Fabry-P\'{e}rot (FP) analysis cavities~\cite{coelho2009three}.

When the pump is weak, no side band is generated. As the pump power increases, more frequency components will emerge~\cite{chembo2010modal}. In this paper, we consider the situation when there are all together five frequency components in the resonator. They are marked as $s2, s1, p, i1, i2$ from lower to higher frequency in the spectrum, as plotted in Fig.~\ref{fig:frequency_conversion} (a). In practice, the total mode number of the comb can be controlled by engineering the device dispesion. In the interaction picture, the Hamiltonian of the system can be written into three parts, i.e.,
\begin{widetext}
\begin{eqnarray}
H=H_{free}+H_{int}+H_{pump}, H_{pump} =  i \hbar \epsilon a_{p}^\dag +H.c. ,\\
H_{int} = i\frac{g}{2} \hbar \sum_{k}a_{k}^\dag a_{k}^\dag a_{k} a_{k}+ ig \hbar \sum_{k \neq t}a_{k}^\dag a_{t}^\dag a_{t} a_{k} \nonumber\\
+ig \hbar(a_{s1}^\dag a_{i1}^\dag a_{p}^2+a_{s2}^\dag a_{i1}^\dag a_{s1} a_{p}+a_{s1}^\dag a_{i2}^\dag a_{i1} a_{p}+a_{s2}^\dag a_{p}^\dag a_{s1}^2 + a_{i2}^\dag a_{p}^\dag a_{i1}^2) +H.c. ,
\end{eqnarray}
\end{widetext}
where $k,t=p,s1,s2,i1,i2$.
The interaction part of the Hamiltonian is derived from the usual Kerr Hamiltonian $V=\hbar (g/2): (a_{p}+a_{s1}+a_{i1}+a_{s2}+a_{i2}+H.c.)^4:$, where ``$:...:$'' stands for normal ordering, $g$ is coupling coefficient which will be explained later, and $\epsilon$ is the pump field that enters the resonator, which, based on non-depletion assumption, is described classically.~\cite{matsko2005optical}
%with application of rotating wave approximation.

The interaction Hamiltonian consists of three parts responsible for self-phase modulation, cross-phase modulation, and four-wave mixing, respectively. It is easy to verify that the first two parts automatically vanish in the P representation~\cite{walls2007quantum}.
We have also omitted some FWM terms indicating the physically less significant nonlinear process for the sake of simplicity, including: {\color{black}
$a_{s2}^\dag a_{i2}^\dag a_{p}^2$, corresponding to the processes that $2p \rightarrow s2 + i2$, is omitted, because the inevitable dispersion will cause farther separated modes be less coupled; and $a_{s2}^\dag a_{i2}^\dag a_{s1} a_{i1}$, corresponding to $s1 + i1 \rightarrow s2 + i2 $, because the absence of pump indicates the process will be much weaker than others. All these processes might actually help the phase-locking mechanism, but are less dominant to the generation of entangled states, so we will leave them for future work.}
In the end there are only five pairs of FWM terms left in the hamiltonian:
$ a_{s1}^\dag a_{i1}^\dag a_{p}^2 $,
$ a_{s2}^\dag a_{p}^\dag a_{s1}^2 $,
$ a_{i2}^\dag a_{p}^\dag a_{i1}^2 $,
$ a_{s2}^\dag a_{i1}^\dag a_{s1} a_{p} $,
$ a_{s1}^\dag a_{i2}^\dag a_{i1} a_{p} $ and their Hermitian counterpart.
The latter two are non-degenerate. Such choice of FWM processes is consistent with the previously mentioned assumptions on the entanglement of multipartite OFCs.

The coupling coefficient $g$ is defined as
\begin{eqnarray}
\label{g}
g=\frac{n_2\hbar\omega_0^2c}{\mathcal{V}n_0^2},
\end{eqnarray}
where $n_2$ is nonlinear refractive index that characterizes the strength of the optical nonlinearity, $n_0$ is the linear refractive index of the material, $c$ is the speed of light in the vacuum.
$\mathcal{V}$ is the mode volume. It can be approximated by $2\pi\lambda R^2$~\cite{matsko2005optical}, yet a more accurate expression is given in ~\cite{chembo2010modal}.
That is to say, if we can expand the intracavity electric field in time domain as
\begin{eqnarray}
\bm{E} (\bm{r},t)=\sum_\mu\frac{1}{2}\varepsilon_\mu (t)\mathrm{e}^{\mathrm{i}\omega_\mu t}\bm{\Upsilon}_\mu (\bm{r})+c.c.,
\end{eqnarray}
where $\varepsilon_\mu (t)$ is the slow-varying envelope. Then each mode volume is calculated as
\begin{eqnarray}
\label{V_mu}
\mathcal{V_\mu}=\left[\int_{V}||\bm{\Upsilon}_\mu (\bm{r})||^4\mathrm{d}V \right]^{-1},
\end{eqnarray}
where $\bm{\Upsilon}_\mu (\bm{r})$ is the normalized eigenvector of the transverse electric (TE) or transverse magnetic (TM) modes.

For a spherical cavity, an analytical solution for $\bm{\Upsilon}_\mu (\bm{r})$ is given in ~\cite{chembo2010modal}. In our case $V_0=6.6\times10^{-12}~\mathrm{m}^3$.
For other cavities, for example, a micro-ring resonator, it might be infeasible to solve analytically~\cite{driscoll2009large}. However, it might still be possible to calculate Eq.\ref{V_mu} numerically utilizing eigenmode solver software such as WGMODES~\cite{fallahkhair2008vector}.

To describe an open system we introduce the loss and out-coupling terms,
\begin{eqnarray}
L_k \rho = \gamma_k (2a_k \rho a_k^\dag - a_k^\dag a_k \rho - \rho a_k^\dag a_k),
\end{eqnarray}
in which $\rho$ is the density matrix of the five cavity modes under consideration.
{\color{black} $\gamma_k = \gamma_{k0}+\gamma_{kc}$ stands for the damping rates for the corresponding resonator modes, which is the sum of two terms. $\gamma_{kc}$ is related to the amplitude transmission coefficients, which is experimentally tunable, and $\gamma_{k0}$ is related to the intracavity absorption rate, determined by the material and shape of the resonator.}

\section{Equations Of Motion For The Full Hamiltonian   \label{sec: Equations Of Motion For The Full Hamiltonian}}
Now consider the whole physical system, where the nonlinear dielectric is contained inside a pumped resonant cavity. The master equation for the first five cavity modes is
\begin{eqnarray}
\frac{\partial\rho}{\partial t}=-\frac{i}{\hbar} [H_{pump}+H_{int},\rho]+\sum_{k=1}^{5}L_k \rho.
\end{eqnarray}
The above master equation can be converted into the equivalent c-number Fockker-Planck equation in P representation,
which may be written as a completely equivalent stochastic differential equation~\cite{walls2007quantum}
\begin{eqnarray}
\label{FP_P}
\frac{\partial\bm{\alpha}}{\partial t}=\bf{F}+\bf{B}\bm{\eta},
\end{eqnarray}
where
$\bm{\alpha} = [\alpha_{p},\alpha_{s1},\alpha_{i1},\alpha_{s2},\alpha_{i2},\alpha_{p}^*,\alpha_{s1}^*,\alpha_{i1}^*,\alpha_{s2}^*,\alpha_{i2}^*]^T$, and $\bm{F}$ is the main part of the system's evolution, in the form of $\bm{F} = [f, f^*]^T$, where
\begin{widetext}
\[ f=\left (  \begin{array}{ccc}
\epsilon-\gamma_{p}\alpha_{p} - 2g \alpha_{p}^*\alpha_{s1}\alpha_{i1} - g \alpha_{s1}^*\alpha_{s2}\alpha_{i1}- g \alpha_{i1}^*\alpha_{i2}\alpha_{s1}  +g \alpha_{s1}^2\alpha_{s2}^* + g \alpha_{i1}^2\alpha_{i2}^*\\
-\gamma_{s1}\alpha_{s1} + g \alpha_{p}^2 \alpha_{i1}^* + g \alpha_{i1}\alpha_{p}\alpha_{i2}^* - g \alpha_{s2}\alpha_{i1}\alpha_{p}^* -2g\alpha_p\alpha_{s2}\alpha_{s1}^*  \\
-\gamma_{i1}\alpha_{i1} + g \alpha_{p}^2 \alpha_{s1}^* + g \alpha_{s1}\alpha_{p}\alpha_{s2}^* - g \alpha_{i2}\alpha_{s1}\alpha_{p}^*  -2g\alpha_p\alpha_{i2}\alpha_{i1}^*\\
-\gamma_{s2}\alpha_{s2} + g \alpha_{s1}\alpha_{p} \alpha_{i1}^* +g\alpha_{s1}^2\alpha_p^*\\
-\gamma_{i2}\alpha_{i2} + g \alpha_{i1}\alpha_{p} \alpha_{s1}^* +g\alpha_{i1}^2\alpha_p^*
\end{array}\right).\]
\end{widetext}
Matrix $\bf{B}$ contains the coefficients of the noise terms thus it is not included in the stability analysis. To obtain $\bf{B}$ we first introduce diffusion matrix
\[ \bf{D}=\left (  \begin{array}{ccc}
\bf{d}  &0\\
0   &\bf{d^*}\end{array}\right),\]
where $d$ is a matrix
\[d=\left (  \begin{array}{ccccc}
-2g\alpha_{s1}\alpha_{i1}    &-g\alpha_{s2}\alpha_{i1}    &-g\alpha_{s1}\alpha_{i2}  &g\alpha_{s1}^2  &g\alpha_{i1}^2\\
-g\alpha_{s2}\alpha_{i1}    &-2g\alpha_{p}\alpha_{s2}    &g\alpha_{p}^2  &0  &g\alpha_{i1}\alpha_{p}\\
-g\alpha_{s1}\alpha_{i2}    &g\alpha_{p}^2  &-2g\alpha_{p}\alpha_{i2}    &g\alpha_{s1}\alpha_{p}   &0\\
g\alpha_{s1}^2   &0  &g\alpha_{s1}\alpha_{p}   &0  &0\\
g\alpha_{i1}^2   &g\alpha_{i1}\alpha_{p}   &0   &0  &0
\end{array}\right).\]
By definition $\bf{B}\bf{B}^T= \bf{D}$, then if the diffusion matrix is positive semi-definite, we can obtain $\bf{B}$ through factoring $\bf{D}$, which is the case in our numerical calculation.
$\bm{\eta} = [\eta_1 (t), \eta_2 (t), \eta_3 (t), \eta_4 (t), \eta_5 (t), c.c.]^T$, where $\eta_i$ are real noise terms characterized by $\left< \eta_i (t) \right>=0$ and $\left< \eta_i (t)\eta_j (t) \right>=\delta_{ij}\delta (t-t')$.

\section{Linearized Quantum-Fluctuation Analysis    \label{sec: Linearized Fluctuation Analysis}}
To solve Eq.\ref{FP_P}, we decompose the system variables into their steady-state (classical) values and negligible quantum fluctuations around the steady-state values as $\alpha_i=A_i+\delta \alpha_i$.
Then one can use linearization analysis as a method to calculate the spectra for the output cavity modes.
The signal photons ($s1$/$s2$) share similar quantum characteristics with their corresponding idlers ($i1$/$i2$). Because $\omega_k$ ($k=p,s1,s2,i1,i2$) are approximately equal, in order to simplify the calculation, we assume all modes have the same intra- and extra-cavity damping rate in the cavity (i.e., $\gamma_{k}=\gamma, \gamma_{kc}=\gamma_c, \gamma_{k0}=\gamma_0, k=p, s1, s2, i1, i2$) in the following part of this paper.

First we calculate the steady-state solution, so $\partial\bm{\alpha}/\partial t$ in Eq.~\ref{FP_P} is set to zero.
The pump threshold is
\begin{eqnarray}
\label{epsilon_th}
\epsilon_{th} = \gamma\sqrt{\gamma/g}.
\end{eqnarray}
When $\epsilon < \epsilon_{th}$, the steady-state outputs can be obtained as $A_{p}=\epsilon/\gamma$ and $A_{j}=0 (i=i1,s1,i2,s2)$. When $\epsilon > \epsilon_{th}$, the steady-state outputs becomes $A_{p}=\frac{\varepsilon+\sqrt{\varepsilon^2+ (3\gamma^3)/g}}{3\gamma}$, $A_{i1}=A_{s1}=A_a=\sqrt{\frac{\gamma}{4g} (1-\frac{\gamma}{gA_p^2})}$ and $A_{i2}=A_{s2}=A_b=\frac{2gA_pA_a^2}{\gamma}$. In the present scheme we only consider the situation for the field modes to oscillate above the threshold. {\color{black} Now that the $\gamma$ is of the order of $10^5$, $A_p=\frac{\varepsilon+\sqrt{\varepsilon^2+3\varepsilon_{th}^2}}{3\gamma}<\varepsilon/\gamma$ is actually much smaller than $\epsilon$, so our non-depletion assumption is self-consistent. }

Then we minus the steady-state part from both sides of the original equation. The fluctuation part is thus linearized as
\begin{eqnarray}
\frac{\partial}{\partial t}\delta\bm{\alpha}=\bm{M}\delta\bm{\alpha}+\bm{B}\bm{\eta},
\end{eqnarray}
where
%$\delta\bm{\alpha}=[\delta\alpha_{p},\delta\alpha_{s1},\delta\alpha_{i1},\delta\alpha_{s2},\delta\alpha_{i2}, \delta\alpha_{p}^*, \delta\alpha_{s1}^*, \delta\alpha_{i1}^*, \delta\alpha_{s2}^*, \delta\alpha_{i2}^*]^T$.
$\delta\bm{\alpha}=[\delta\alpha_{p},\delta\alpha_{s1},\delta\alpha_{i1},\delta\alpha_{s2},\delta\alpha_{i2}, H.c.]^T$.
M is the drift matrix with the steady-state values inserted, given by
\begin{widetext}
\[\bm{M}=\left (  \begin{array}{cc}
\bm{m}_1 &\bm{m}_2\\
\bm{m}_2^* &\bm{m}_1^*
\end{array}\right),\]

\[\bm{m}_1=\left (  \begin{array}{ccccc}
-\gamma           &-gA_aA_b-2gA_pA_a+2gA_aA_b   &-gA_aA_b-2gA_pA_a+2gA_aA_b   &-gA_a^2    &-gA_a^2\\
gA_aA_b+2gA_pA_a-2gA_aA_b    &-\gamma          &0                  &-3gA_pA_a   &0\\
gA_aA_b+2gA_pA_a-2gA_aA_b    &0                  &-\gamma          &0          &-3gA_pA_a\\
gA_a^2              &3gA_pA_a            &0                    &-\gamma  &0\\
gA_a^2              &0                   &3gA_pA_a           &0            &-\gamma
\end{array}\right),\]

\[\bm{m}_2=\left (  \begin{array}{ccccc}
-2g\alpha_{a}^2    &-g\alpha_{a}\alpha_{b}    &-g\alpha_{a}\alpha_{b}  &g\alpha_{a}^2  &g\alpha_{a}^2\\
-g\alpha_{a}\alpha_{b}    &-2g\alpha_{p}\alpha_{b}    &g\alpha_{p}^2  &0  &g\alpha_{a}\alpha_{p}\\
-g\alpha_{a}\alpha_{b}    &g\alpha_{p}^2  &-2g\alpha_{p}\alpha_{b}    &g\alpha_{a}\alpha_{p}   &0\\
g\alpha_{a}^2   &0  &g\alpha_{a}\alpha_{p}   &0  &0\\
g\alpha_{a}^2   &g\alpha_{a}\alpha_{p}   &0   &0  &0
\end{array}\right).\]

\end{widetext}
For the linearized quantum-fluctuation analysis to be valid the fluctuations must remain small compared to the mean values and the eigenvalues of the drift matrix $\bm{M}$ must have no negative real part. If the requirement that the real part of the eigenvalues of $-\bm{M}$ stay positive is satisfied, the fluctuation equations will describe an Ornstein-Uhlenbeck process~\cite{gardiner1985stochastic} for which the intracavity spectral correlation matrix is
\begin{eqnarray}
\label{S}
\bm{S} (\omega)= (-\bm{M}+\mathrm{i}\omega\bm{I})^{-1}\bm{D} (-\bm{M}^T-\mathrm{i}\omega\bm{I})^{-1}.
\end{eqnarray}
All the correlations required to study the measurable extracavity spectra are contained in this intracavity spectral matrix.

In order to investigate multipartite entanglement, and to show that the system under consideration demonstrates effective five-partite entanglement, we first define quadrature operators  for each mode as
\begin{eqnarray}
X_k = a_k + a_k^\dag, Y_k = -i (a_k - a_k^\dag),
\end{eqnarray}
therefore $[X_k, Y_k]=2i$.
%Based on such definition, $V (X_k) \leq 1$ will indicate a squeezed state.
$V (A)=\left<A^2 \right>- \left< A \right>^2$ denotes the variance of operator $A$.

The output fields is determined by the well-known input-output relations~\cite{gardiner1985input}. In particular, the spectral variances and covariances have the general form
\begin{eqnarray}
\label{input_output}
S_{X_i}^{out} (\omega)=1+2\gamma_{c} S_{X_i} (\omega)\\
S_{X_i,X_j}^{out} (\omega)=2\gamma_{c} S_{X_i,X_j} (\omega)
\end{eqnarray}
Similar expressions can be derived for the $Y$ quadratures.

\section{Five-partite Entanglement Criteria    \label{sec: Five-partite Entanglement Measures}}
The condition proposed by van Loock and Furusawa~\cite{loock2003detecting}, which is a generalization of the conditions for bipartite entanglement, is sufficient to demonstrate multipartite entanglement. We now demonstrate how these may be optimized for the verification of genuine five-partite entanglement in this system.
Using the quadrature definitions, the five-partite inequalities, which must be simultaneously violated, are
\begin{widetext}
\begin{eqnarray}
\label{X_p+X_s1}
S_{ (1)} = V (X_{p}+X_{s1}) + V (-Y_{p} + Y_{s1} + g_{i1}Y_{i1} + g_{s2}Y_{s2} + g_{i2}Y_{i2}) \geq 4,\\
\label{X_p+X_i1}
S_{ (2)} = V (X_{p}+X_{i1}) + V (-Y_{p} + g_{s1}Y_{s1} + Y_{i1} + g_{s2}Y_{s2} + g_{i2}Y_{i2}) \geq 4,\\
\label{X_s1-X_s2}
S_{ (3)} = V (X_{s1}-X_{s2})+ V (g_{p}Y_{p} + Y_{s1} + g_{i1}Y_{i1} + Y_{s2} + g_{i2}Y_{i2}) \geq 4,\\
\label{X_i2-X_i1}
S_{ (4)} = V (X_{i2}-X_{i1})+ V (g_{p}Y_{p} + g_{s1}Y_{s1} + Y_{i1} + g_{s2}Y_{s2} + Y_{i2}) \geq 4,
\end{eqnarray}
\end{widetext}
where the $g_k (k=p,s1,s2,i1,i2)$ are arbitrary real parameters that are used to optimize the violation of these inequalities. It is important to note that in the uncorrelated limit these optimized
VLF criterion approach 4. Hence, without optimization, some entanglement which is present may be missed.

\section{Output Fluctuation Spectra \label{sec: Output Fluctuation Spectra}}
From now on, we numerically calculate the values of VLF inequalities according to the results obtained above.
We assume that the resonator is a spherical $\mathrm{CaF}_2$ cavity. Note that the theoretical model and analysis are universal and can be easily applied to other device platforms. For ${\rm CaF_{2}}$, the refractive index is $n_0=1.43$, Kerr coefficient is $n_2=3.2\times10^{-20}~\mathrm{m^2/W}$. We assume the $\mathrm{CaF}_2$ resonator has a radius $R$ of $2.5~\mathrm{mm}$ and light is critically coupled to the device with a loaded quality factor $Q_0=3\times10^9$ (corresponding to a central modal bandwidth $\Delta \omega_0 =\gamma_{p} \approx 2\pi\times64~\mathrm{kHz}$~\cite{chembo2010modal}), with a pump launched at a wavelength of $\lambda_0=1560.5~\mathrm{nm}$.

%Because $\omega_k$ ($k=p,s1,s2,i1,i2$) are approximately equal, we assume $\gamma_a=\gamma_b=\gamma_p$ in the following analysis.
Due to the symmetry relation between signal and idler photons, Eq.\ref{X_p+X_s1} and Eq.\ref{X_p+X_i1} are equivalent. So are Eq.\ref{X_s1-X_s2} and Eq.\ref{X_i2-X_i1}. Therefore, we only need to calculate $S_{ (1)}$ and $S_{ (3)}$.

\begin{figure*}
\includegraphics[width=16cm]{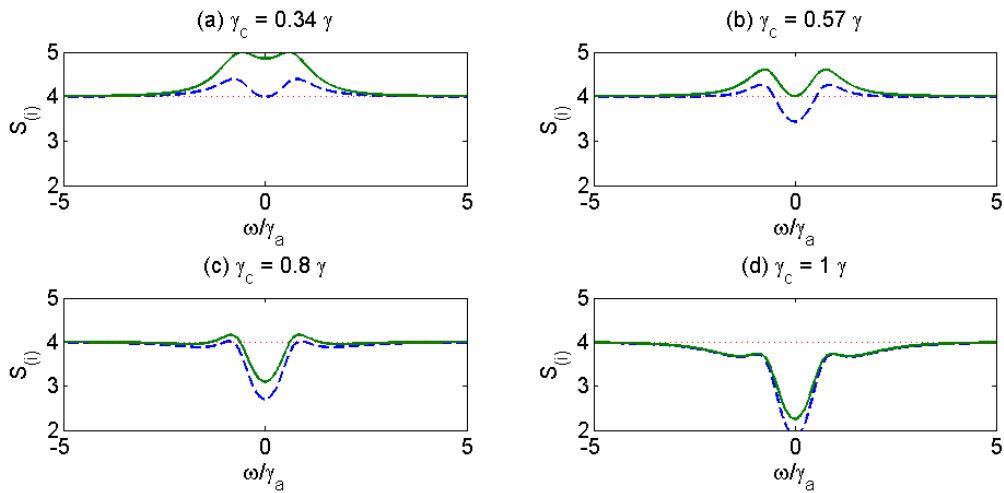}% Here is how to import EPS art
\caption{\label{fig:4in1}Four variance versus frequency of pump plots when $\gamma_{c}$ is $0.34, 0.57, 0.8, 1$ times as great as $\gamma$ (from left to right and top to bottom). {\color{black} $\gamma=4.02\times10^5 \mathrm{s}^{-1}$, $g=2.21\times10^{-4} \mathrm{s}^{-1}$, $\epsilon = 1.15\epsilon_{th} = 1.97\times10^{10} \mathrm{s}^{-1}$}.
The blue dashed curve stand for $S_{ (1)}$ and $S_{ (2)}$, whereas the green solid ones stand for $S_{ (3)}$ and $S_{ (4)}$. The pump power is fixed at $1.15\epsilon_{th}$.
}
\end{figure*}

To begin with, it is evident from the Eq.~\ref{FP_P} through \ref{input_output} that the intracavity entanglement is completely determined once total damping rate $\gamma$, coupling coefficient $g$ and pumping power $\epsilon$ is fixed, but it is always the variables outside the cavity that we observe, so the transfer also plays a role in the observation, which is mathematically determined by a ratio, $\gamma_{c}/\gamma$. Thus we first fix the first three factors and vary the ratio to see how it affect the observed entanglement. In Fig.\ref{fig:4in1}, we plot the minimum of the variances versus the analysis frequency normalized to $\gamma$ when $\gamma_{c}$ takes a portion of $0.34, 0.57, 0.8, 1$ of the total damping rate. The blue dashed lines stand for $S_{ (1)}$, while the green solid ones stand for $S_{ (3)}$.

It can be inferred from the plot that when $\gamma_{c}=0.34\gamma$, there is no entanglement between any two of the field modes. As we increase the out-coupling coefficients, the $s1$ and $i1$ begin to entangle with the pump photons around the center frequency, but it is not until when $\gamma_{c}/\gamma=0.57$ that the $s2$ and $i2$ begin to entangle with $s1$ and $i1$, respectively. Eventually the variance converge to Fig.\ref{fig:4in1} (d).
Thus we conclude that the entanglement among output modes are improved as the $\gamma_{c}/\gamma$ ratio increase, i.e., the entanglement is better when the cavity has higher $Q$ therefore lower intracavity loss, and higher extracavity coupling coefficient. This can be interpreted naturally if we see the coupling as a beam splitter which extract squeezed quantum noise to the output\cite{gardiner1985input}, so the higher portion the coupling coefficient takes in the total damping rate, the less consumed entangled pair of photons are wasted in the internal loss. For that consideration, we will ideally fix $\gamma_{0}=0$ in the following analysis, so that the effect of output transfer is suppressed to minimum.

\begin{figure}
\includegraphics[width=8cm]{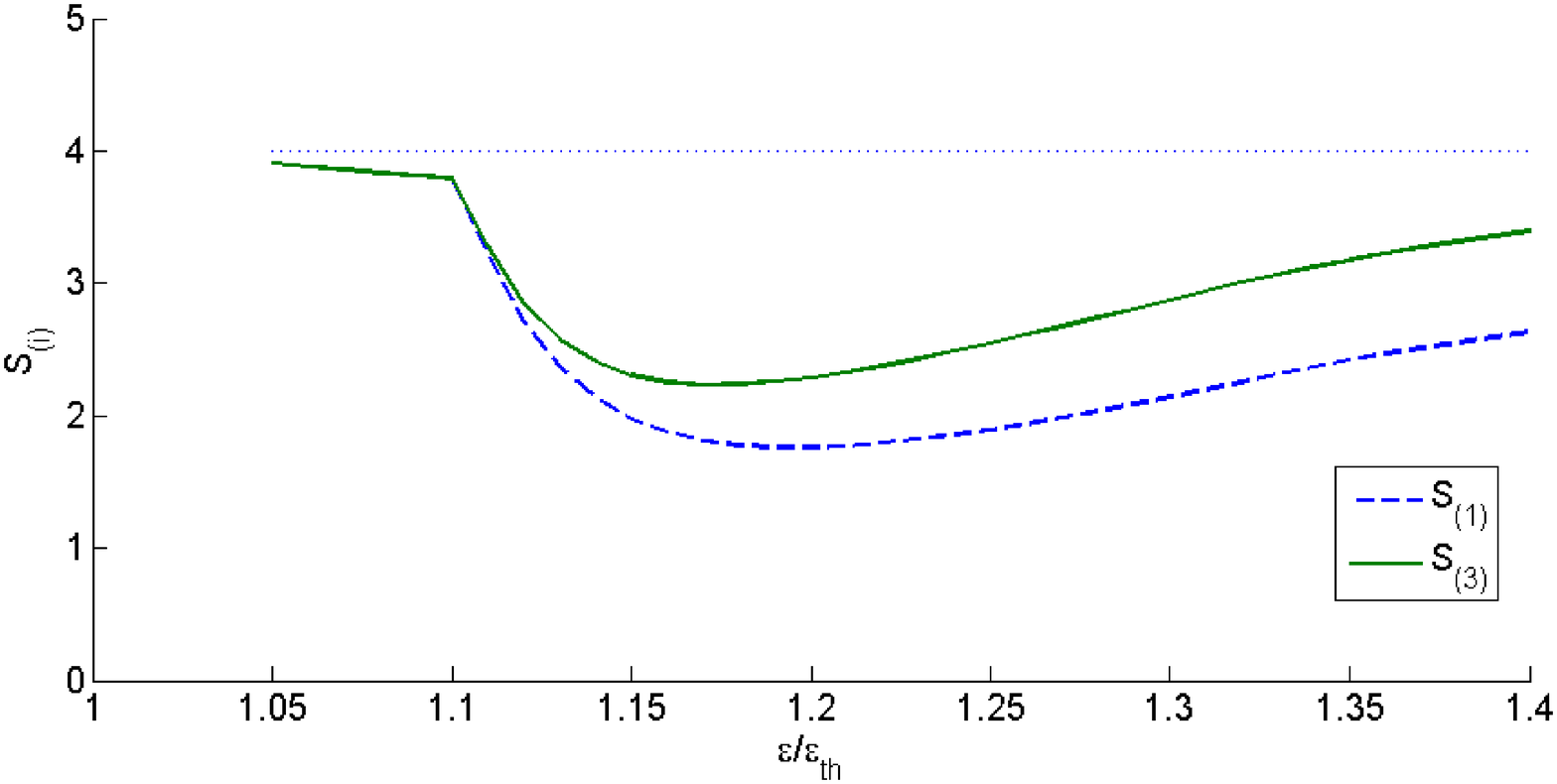}% Here is how to import EPS art
\caption{\label{fig:varianceVSpump}The minimum variance as a function of pump power.
%The blue dashed curve stand for $S_{ (1)}$ and $S_{ (2)}$, whereas the green solid ones stand for that of $S_{ (3)}$ and $S_{ (4)}$.
}
\end{figure}

\begin{figure*}
\includegraphics[width=16cm]{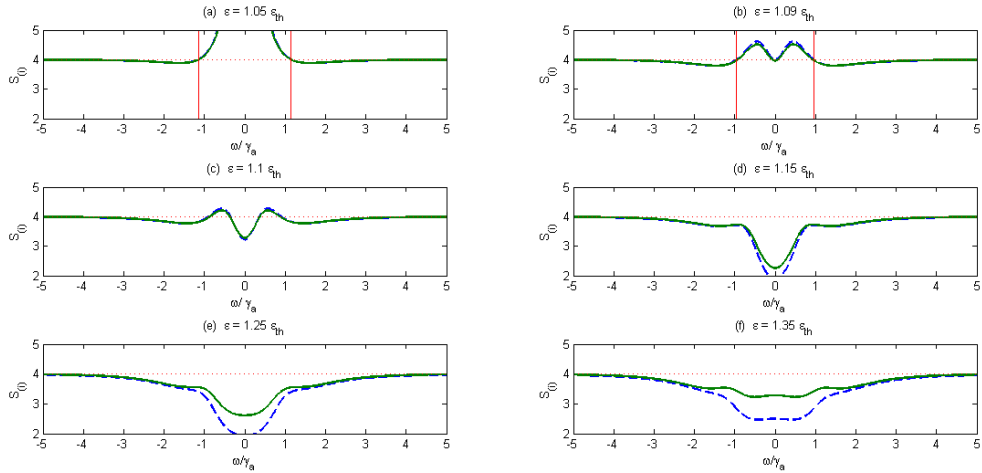}% Here is how to import EPS art
\caption{\label{fig:6in1}The variance versus frequency under different pumping power.
%The blue dashed curve stand for $S_{ (1)}$ and $S_{ (2)}$, whereas the green solid ones stand for that of $S_{ (3)}$ and $S_{ (4)}$.
}
\end{figure*}

The pump power, on the other hand, is more subtle. It can be derived from our previous analysis that once $\epsilon/\epsilon_{th}$ and $\omega/\gamma$ is fixed, the variance is totally determined. In other words, the variance $S_{i}$ as a function of $\omega/\gamma$ is solely determined by the parameter $\epsilon/\epsilon_{th}$ instead of $g$, $\gamma$, or $\varepsilon$ independently. This might come counter-intuitive in the first sight, as we are indicating that the noise spectrum will look just the same except for a scaling in the frequency axis if we the half $n_2$ and $\gamma$ simultaneously, as long as we keep $\epsilon/\epsilon_{th}$ constant. {\color{black} However, we should also notice that the bandwidth $\Delta\omega$ when five-partite entanglement is witnessed will also be halved, indicating a longer entangled period in the time domain. }
When the pump is too weak (less than the threshold), there will be no sideband, at least not all of them. If the pump grows too strong, additional sideband that we didn't take into consideration might pop up, deteriorating the effectiveness of our model. We plot the minimal variance throughout the noise power spectrum as a function of the pump power (normalized by $\epsilon_{\mathrm{th}}$) in Fig.~\ref{fig:varianceVSpump} and six typical spectrums in Fig.~\ref{fig:6in1}.
%{\color{black} Why rose again?}

%Again, The blue dashed curve stand for $S_{ (1)}$ and $S_{ (2)}$, whereas the green solid one stand for $S_{ (3)}$ and $S_{ (4)}$.
It can be inferred from the graphs that both variance first descend as the pump power power increases, then ascend. $S_{ (3)}$ and $S_{ (4)}$ reaches their global minimum at $\epsilon=1.15\epsilon_{\mathrm{th}}$. Considering that they are the short slabs of the whole entanglement system, we conclude that $1.15\epsilon_{\mathrm{th}}$ is the optimal pump power. The other turning point in Fig.~\ref{fig:varianceVSpump} is around $1.1\epsilon_{\mathrm{th}}$, when, as we can see in Fig.~\ref{fig:6in1}~(b) and (c), the variances in the center frequency begin to decrease dramatically and become the minimum which was once achieved in the side band, as showed in Fig.~\ref{fig:6in1}~(a).

\section{Conclusions \label{sec: Discussion and conclusions}}

In conclusion, we presented a theoretical model for the five-partite continuous-variable entanglement among five field modes based on cascaded four wave mixing process.
By solving Fokker-Planck equation in P representation, we analyzed the entanglement condition when van Loock and Furusawa criteria are violated.
We presented the design parameters for experimental purpose, and they might also be utilized to build integrated compact five-partite entanglement generator.
We found that the degree of entanglement was totally determined by $\omega/\gamma$, $\epsilon/\epsilon_{th}$ and $\gamma_c/\gamma$.
This result filled the blank in theory for the entanglement analysis of OFCs generated from high-Q resonator, therefore would pave the way for future optical quantum computation on chip.

\section{Acknowledgement}
This work is supported by the National Natural Science Foundation of China (Grant Nos. 61475099, 61102053), Program of State Key Laboratory of Quantum Optics and Quantum Optics Devices (No£ºKF201405), the Scientific Research Foundation for the Returned Overseas Scholars, State Education Ministry, SMC Excellent Young Faculty program, SJTU 2011. Q.L also thanks the support of National Science Foundation of the United States under Grant No. ECCS-1351697.

\bibliography{ref}% Produces the bibliography via BibTeX.

\begin{thebibliography}{37}%
\makeatletter
\providecommand \@ifxundefined [1]{%
 \@ifx{#1\undefined}
}%
\providecommand \@ifnum [1]{%
 \ifnum #1\expandafter \@firstoftwo
 \else \expandafter \@secondoftwo
 \fi
}%
\providecommand \@ifx [1]{%
 \ifx #1\expandafter \@firstoftwo
 \else \expandafter \@secondoftwo
 \fi
}%
\providecommand \natexlab [1]{#1}%
\providecommand \enquote  [1]{``#1''}%
\providecommand \bibnamefont  [1]{#1}%
\providecommand \bibfnamefont [1]{#1}%
\providecommand \citenamefont [1]{#1}%
\providecommand \href@noop [0]{\@secondoftwo}%
\providecommand \href [0]{\begingroup \@sanitize@url \@href}%
\providecommand \@href[1]{\@@startlink{#1}\@@href}%
\providecommand \@@href[1]{\endgroup#1\@@endlink}%
\providecommand \@sanitize@url [0]{\catcode `\\12\catcode `\$12\catcode
  `\&12\catcode `\#12\catcode `\^12\catcode `\_12\catcode `\%12\relax}%
\providecommand \@@startlink[1]{}%
\providecommand \@@endlink[0]{}%
\providecommand \url  [0]{\begingroup\@sanitize@url \@url }%
\providecommand \@url [1]{\endgroup\@href {#1}{\urlprefix }}%
\providecommand \urlprefix  [0]{URL }%
\providecommand \Eprint [0]{\href }%
\providecommand \doibase [0]{http://dx.doi.org/}%
\providecommand \selectlanguage [0]{\@gobble}%
\providecommand \bibinfo  [0]{\@secondoftwo}%
\providecommand \bibfield  [0]{\@secondoftwo}%
\providecommand \translation [1]{[#1]}%
\providecommand \BibitemOpen [0]{}%
\providecommand \bibitemStop [0]{}%
\providecommand \bibitemNoStop [0]{.\EOS\space}%
\providecommand \EOS [0]{\spacefactor3000\relax}%
\providecommand \BibitemShut  [1]{\csname bibitem#1\endcsname}%
\let\auto@bib@innerbib\@empty
%</preamble>
\bibitem [{\citenamefont {Shor}(1997)}]{shor1997polynomial}%
  \BibitemOpen
  \bibfield  {author} {\bibinfo {author} {\bibfnamefont {P.~W.}\ \bibnamefont
  {Shor}},\ }\href@noop {} {\bibfield  {journal} {\bibinfo  {journal} {SIAM
  Journal on Computing}\ }\textbf {\bibinfo {volume} {26}},\ \bibinfo {pages}
  {1484} (\bibinfo {year} {1997})}\BibitemShut {NoStop}%
\bibitem [{\citenamefont {Feynman}(1982)}]{feynman1982simulating}%
  \BibitemOpen
  \bibfield  {author} {\bibinfo {author} {\bibfnamefont {R.~P.}\ \bibnamefont
  {Feynman}},\ }\href@noop {} {\bibfield  {journal} {\bibinfo  {journal}
  {International Journal of Theoretical Physics}\ }\textbf {\bibinfo {volume}
  {21}},\ \bibinfo {pages} {467} (\bibinfo {year} {1982})}\BibitemShut
  {NoStop}%
\bibitem [{\citenamefont {Raussendorf}\ and\ \citenamefont
  {Briegel}(2001)}]{raussendorf2001one}%
  \BibitemOpen
  \bibfield  {author} {\bibinfo {author} {\bibfnamefont {R.}~\bibnamefont
  {Raussendorf}}\ and\ \bibinfo {author} {\bibfnamefont {H.~J.}\ \bibnamefont
  {Briegel}},\ }\href@noop {} {\bibfield  {journal} {\bibinfo  {journal}
  {Physical Review Letters}\ }\textbf {\bibinfo {volume} {86}},\ \bibinfo
  {pages} {5188} (\bibinfo {year} {2001})}\BibitemShut {NoStop}%
\bibitem [{\citenamefont {Briegel}\ and\ \citenamefont
  {Raussendorf}(2001)}]{briegel2001persistent}%
  \BibitemOpen
  \bibfield  {author} {\bibinfo {author} {\bibfnamefont {H.~J.}\ \bibnamefont
  {Briegel}}\ and\ \bibinfo {author} {\bibfnamefont {R.}~\bibnamefont
  {Raussendorf}},\ }\href@noop {} {\bibfield  {journal} {\bibinfo  {journal}
  {Physical Review Letters}\ }\textbf {\bibinfo {volume} {86}},\ \bibinfo
  {pages} {910} (\bibinfo {year} {2001})}\BibitemShut {NoStop}%
\bibitem [{\citenamefont {Menicucci}\ \emph {et~al.}(2008)\citenamefont
  {Menicucci}, \citenamefont {Flammia},\ and\ \citenamefont
  {Pfister}}]{menicucci2008one}%
  \BibitemOpen
  \bibfield  {author} {\bibinfo {author} {\bibfnamefont {N.~C.}\ \bibnamefont
  {Menicucci}}, \bibinfo {author} {\bibfnamefont {S.~T.}\ \bibnamefont
  {Flammia}}, \ and\ \bibinfo {author} {\bibfnamefont {O.}~\bibnamefont
  {Pfister}},\ }\href@noop {} {\bibfield  {journal} {\bibinfo  {journal}
  {Physical Review Letters}\ }\textbf {\bibinfo {volume} {101}},\ \bibinfo
  {pages} {130501} (\bibinfo {year} {2008})}\BibitemShut {NoStop}%
\bibitem [{\citenamefont {Roslund}\ \emph {et~al.}(2013)\citenamefont
  {Roslund}, \citenamefont {De~Araujo}, \citenamefont {Jiang}, \citenamefont
  {Fabre},\ and\ \citenamefont {Treps}}]{roslund2013wavelength}%
  \BibitemOpen
  \bibfield  {author} {\bibinfo {author} {\bibfnamefont {J.}~\bibnamefont
  {Roslund}}, \bibinfo {author} {\bibfnamefont {R.~M.}\ \bibnamefont
  {De~Araujo}}, \bibinfo {author} {\bibfnamefont {S.}~\bibnamefont {Jiang}},
  \bibinfo {author} {\bibfnamefont {C.}~\bibnamefont {Fabre}}, \ and\ \bibinfo
  {author} {\bibfnamefont {N.}~\bibnamefont {Treps}},\ }\href@noop {}
  {\bibfield  {journal} {\bibinfo  {journal} {Nature Photonics}\ } (\bibinfo
  {year} {2013})}\BibitemShut {NoStop}%
\bibitem [{\citenamefont {Cundiff}\ and\ \citenamefont {Ye}(2003)}]{Cundiff03}%
  \BibitemOpen
  \bibfield  {author} {\bibinfo {author} {\bibfnamefont {S.~T.}\ \bibnamefont
  {Cundiff}}\ and\ \bibinfo {author} {\bibfnamefont {J.}~\bibnamefont {Ye}},\
  }\href@noop {} {\bibfield  {journal} {\bibinfo  {journal} {Reviews of Modern
  Physics}\ }\textbf {\bibinfo {volume} {75}},\ \bibinfo {pages} {325}
  (\bibinfo {year} {2003})}\BibitemShut {NoStop}%
\bibitem [{\citenamefont {Diddams}(2010)}]{Diddams10}%
  \BibitemOpen
  \bibfield  {author} {\bibinfo {author} {\bibfnamefont {S.~A.}\ \bibnamefont
  {Diddams}},\ }\href@noop {} {\bibfield  {journal} {\bibinfo  {journal} {JOSA
  B}\ }\textbf {\bibinfo {volume} {27}},\ \bibinfo {pages} {B51} (\bibinfo
  {year} {2010})}\BibitemShut {NoStop}%
\bibitem [{\citenamefont {Del¡¯Haye}\ \emph {et~al.}(2007)\citenamefont
  {Del¡¯Haye}, \citenamefont {Schliesser}, \citenamefont {Arcizet},
  \citenamefont {Wilken}, \citenamefont {Holzwarth},\ and\ \citenamefont
  {Kippenberg}}]{del2007optical}%
  \BibitemOpen
  \bibfield  {author} {\bibinfo {author} {\bibfnamefont {P.}~\bibnamefont
  {Del¡¯Haye}}, \bibinfo {author} {\bibfnamefont {A.}~\bibnamefont
  {Schliesser}}, \bibinfo {author} {\bibfnamefont {O.}~\bibnamefont {Arcizet}},
  \bibinfo {author} {\bibfnamefont {T.}~\bibnamefont {Wilken}}, \bibinfo
  {author} {\bibfnamefont {R.}~\bibnamefont {Holzwarth}}, \ and\ \bibinfo
  {author} {\bibfnamefont {T.}~\bibnamefont {Kippenberg}},\ }\href@noop {}
  {\bibfield  {journal} {\bibinfo  {journal} {Nature}\ }\textbf {\bibinfo
  {volume} {450}},\ \bibinfo {pages} {1214} (\bibinfo {year}
  {2007})}\BibitemShut {NoStop}%
\bibitem [{\citenamefont {Kippenberg}\ \emph {et~al.}(2011)\citenamefont
  {Kippenberg}, \citenamefont {Holzwarth},\ and\ \citenamefont
  {Diddams}}]{kippenberg2011microresonator}%
  \BibitemOpen
  \bibfield  {author} {\bibinfo {author} {\bibfnamefont {T.}~\bibnamefont
  {Kippenberg}}, \bibinfo {author} {\bibfnamefont {R.}~\bibnamefont
  {Holzwarth}}, \ and\ \bibinfo {author} {\bibfnamefont {S.}~\bibnamefont
  {Diddams}},\ }\href@noop {} {\bibfield  {journal} {\bibinfo  {journal}
  {Science}\ }\textbf {\bibinfo {volume} {332}},\ \bibinfo {pages} {555}
  (\bibinfo {year} {2011})}\BibitemShut {NoStop}%
\bibitem [{\citenamefont {Chembo}\ and\ \citenamefont
  {Yu}(2010)}]{chembo2010modal}%
  \BibitemOpen
  \bibfield  {author} {\bibinfo {author} {\bibfnamefont {Y.~K.}\ \bibnamefont
  {Chembo}}\ and\ \bibinfo {author} {\bibfnamefont {N.}~\bibnamefont {Yu}},\
  }\href@noop {} {\bibfield  {journal} {\bibinfo  {journal} {Physical Review
  A}\ }\textbf {\bibinfo {volume} {82}},\ \bibinfo {pages} {033801} (\bibinfo
  {year} {2010})}\BibitemShut {NoStop}%
\bibitem [{\citenamefont {Walls}\ and\ \citenamefont
  {Milburn}(2007)}]{walls2007quantum}%
  \BibitemOpen
  \bibfield  {author} {\bibinfo {author} {\bibfnamefont {D.~F.}\ \bibnamefont
  {Walls}}\ and\ \bibinfo {author} {\bibfnamefont {G.~J.}\ \bibnamefont
  {Milburn}},\ }\href@noop {} {\emph {\bibinfo {title} {Quantum Optics}}}\
  (\bibinfo  {publisher} {Springer},\ \bibinfo {year} {2007})\BibitemShut
  {NoStop}%
\bibitem [{\citenamefont {Shelby}\ \emph {et~al.}(1986)\citenamefont {Shelby},
  \citenamefont {Levenson}, \citenamefont {Perlmutter}, \citenamefont {DeVoe},\
  and\ \citenamefont {Walls}}]{Shelby86}%
  \BibitemOpen
  \bibfield  {author} {\bibinfo {author} {\bibfnamefont {R.~M.}\ \bibnamefont
  {Shelby}}, \bibinfo {author} {\bibfnamefont {M.~D.}\ \bibnamefont
  {Levenson}}, \bibinfo {author} {\bibfnamefont {S.~H.}\ \bibnamefont
  {Perlmutter}}, \bibinfo {author} {\bibfnamefont {R.~G.}\ \bibnamefont
  {DeVoe}}, \ and\ \bibinfo {author} {\bibfnamefont {D.~F.}\ \bibnamefont
  {Walls}},\ }\href@noop {} {\bibfield  {journal} {\bibinfo  {journal}
  {Physical review letters}\ }\textbf {\bibinfo {volume} {57}},\ \bibinfo
  {pages} {691} (\bibinfo {year} {1986})}\BibitemShut {NoStop}%
\bibitem [{\citenamefont {Wu}\ \emph {et~al.}(1986)\citenamefont {Wu},
  \citenamefont {Kimble}, \citenamefont {Hall},\ and\ \citenamefont
  {Wu}}]{Kimble86}%
  \BibitemOpen
  \bibfield  {author} {\bibinfo {author} {\bibfnamefont {L.-A.}\ \bibnamefont
  {Wu}}, \bibinfo {author} {\bibfnamefont {H.~J.}\ \bibnamefont {Kimble}},
  \bibinfo {author} {\bibfnamefont {J.~L.}\ \bibnamefont {Hall}}, \ and\
  \bibinfo {author} {\bibfnamefont {H.}~\bibnamefont {Wu}},\ }\href@noop {}
  {\bibfield  {journal} {\bibinfo  {journal} {Physical review letters}\
  }\textbf {\bibinfo {volume} {57}},\ \bibinfo {pages} {2520} (\bibinfo {year}
  {1986})}\BibitemShut {NoStop}%
\bibitem [{\citenamefont {Wu}\ \emph {et~al.}(1987)\citenamefont {Wu},
  \citenamefont {Xiao},\ and\ \citenamefont {Kimble}}]{Kimble87}%
  \BibitemOpen
  \bibfield  {author} {\bibinfo {author} {\bibfnamefont {L.-A.}\ \bibnamefont
  {Wu}}, \bibinfo {author} {\bibfnamefont {M.}~\bibnamefont {Xiao}}, \ and\
  \bibinfo {author} {\bibfnamefont {H.}~\bibnamefont {Kimble}},\ }\href@noop {}
  {\bibfield  {journal} {\bibinfo  {journal} {JOSA B}\ }\textbf {\bibinfo
  {volume} {4}},\ \bibinfo {pages} {1465} (\bibinfo {year} {1987})}\BibitemShut
  {NoStop}%
\bibitem [{\citenamefont {Wolinsky}\ and\ \citenamefont
  {Carmichael}(1988)}]{Carmichael88}%
  \BibitemOpen
  \bibfield  {author} {\bibinfo {author} {\bibfnamefont {M.}~\bibnamefont
  {Wolinsky}}\ and\ \bibinfo {author} {\bibfnamefont {H.~J.}\ \bibnamefont
  {Carmichael}},\ }\href@noop {} {\bibfield  {journal} {\bibinfo  {journal}
  {Physical review letters}\ }\textbf {\bibinfo {volume} {60}},\ \bibinfo
  {pages} {1836} (\bibinfo {year} {1988})}\BibitemShut {NoStop}%
\bibitem [{\citenamefont {Breitenbach}\ \emph {et~al.}(1995)\citenamefont
  {Breitenbach}, \citenamefont {M{\"u}ller}, \citenamefont {Pereira},
  \citenamefont {Poizat}, \citenamefont {Schiller},\ and\ \citenamefont
  {Mlynek}}]{Mlynek95}%
  \BibitemOpen
  \bibfield  {author} {\bibinfo {author} {\bibfnamefont {G.}~\bibnamefont
  {Breitenbach}}, \bibinfo {author} {\bibfnamefont {T.}~\bibnamefont
  {M{\"u}ller}}, \bibinfo {author} {\bibfnamefont {S.}~\bibnamefont {Pereira}},
  \bibinfo {author} {\bibfnamefont {J.-P.}\ \bibnamefont {Poizat}}, \bibinfo
  {author} {\bibfnamefont {S.}~\bibnamefont {Schiller}}, \ and\ \bibinfo
  {author} {\bibfnamefont {J.}~\bibnamefont {Mlynek}},\ }\href@noop {}
  {\bibfield  {journal} {\bibinfo  {journal} {JOSA B}\ }\textbf {\bibinfo
  {volume} {12}},\ \bibinfo {pages} {2304} (\bibinfo {year}
  {1995})}\BibitemShut {NoStop}%
\bibitem [{\citenamefont {Breitenbach}\ \emph {et~al.}(1997)\citenamefont
  {Breitenbach}, \citenamefont {Schiller},\ and\ \citenamefont
  {Mlynek}}]{Mlynek97}%
  \BibitemOpen
  \bibfield  {author} {\bibinfo {author} {\bibfnamefont {G.}~\bibnamefont
  {Breitenbach}}, \bibinfo {author} {\bibfnamefont {S.}~\bibnamefont
  {Schiller}}, \ and\ \bibinfo {author} {\bibfnamefont {J.}~\bibnamefont
  {Mlynek}},\ }\href@noop {} {\bibfield  {journal} {\bibinfo  {journal}
  {Nature}\ }\textbf {\bibinfo {volume} {387}},\ \bibinfo {pages} {471}
  (\bibinfo {year} {1997})}\BibitemShut {NoStop}%
\bibitem [{\citenamefont {Laurat}\ \emph {et~al.}(2005)\citenamefont {Laurat},
  \citenamefont {Longchambon}, \citenamefont {Fabre},\ and\ \citenamefont
  {Coudreau}}]{Laurat05}%
  \BibitemOpen
  \bibfield  {author} {\bibinfo {author} {\bibfnamefont {J.}~\bibnamefont
  {Laurat}}, \bibinfo {author} {\bibfnamefont {L.}~\bibnamefont {Longchambon}},
  \bibinfo {author} {\bibfnamefont {C.}~\bibnamefont {Fabre}}, \ and\ \bibinfo
  {author} {\bibfnamefont {T.}~\bibnamefont {Coudreau}},\ }\href@noop {}
  {\bibfield  {journal} {\bibinfo  {journal} {Optics letters}\ }\textbf
  {\bibinfo {volume} {30}},\ \bibinfo {pages} {1177} (\bibinfo {year}
  {2005})}\BibitemShut {NoStop}%
\bibitem [{\citenamefont {Vahlbruch}\ \emph {et~al.}(2008)\citenamefont
  {Vahlbruch}, \citenamefont {Mehmet}, \citenamefont {Chelkowski},
  \citenamefont {Hage}, \citenamefont {Franzen}, \citenamefont {Lastzka},
  \citenamefont {Go{\ss}ler}, \citenamefont {Danzmann},\ and\ \citenamefont
  {Schnabel}}]{Vahlbruch08}%
  \BibitemOpen
  \bibfield  {author} {\bibinfo {author} {\bibfnamefont {H.}~\bibnamefont
  {Vahlbruch}}, \bibinfo {author} {\bibfnamefont {M.}~\bibnamefont {Mehmet}},
  \bibinfo {author} {\bibfnamefont {S.}~\bibnamefont {Chelkowski}}, \bibinfo
  {author} {\bibfnamefont {B.}~\bibnamefont {Hage}}, \bibinfo {author}
  {\bibfnamefont {A.}~\bibnamefont {Franzen}}, \bibinfo {author} {\bibfnamefont
  {N.}~\bibnamefont {Lastzka}}, \bibinfo {author} {\bibfnamefont
  {S.}~\bibnamefont {Go{\ss}ler}}, \bibinfo {author} {\bibfnamefont
  {K.}~\bibnamefont {Danzmann}}, \ and\ \bibinfo {author} {\bibfnamefont
  {R.}~\bibnamefont {Schnabel}},\ }\href@noop {} {\bibfield  {journal}
  {\bibinfo  {journal} {Physical review letters}\ }\textbf {\bibinfo {volume}
  {100}},\ \bibinfo {pages} {033602} (\bibinfo {year} {2008})}\BibitemShut
  {NoStop}%
\bibitem [{\citenamefont {Yonezawa}\ \emph {et~al.}(2010)\citenamefont
  {Yonezawa}, \citenamefont {Nagashima},\ and\ \citenamefont
  {Furusawa}}]{Furusawa10}%
  \BibitemOpen
  \bibfield  {author} {\bibinfo {author} {\bibfnamefont {H.}~\bibnamefont
  {Yonezawa}}, \bibinfo {author} {\bibfnamefont {K.}~\bibnamefont {Nagashima}},
  \ and\ \bibinfo {author} {\bibfnamefont {A.}~\bibnamefont {Furusawa}},\
  }\href@noop {} {\bibfield  {journal} {\bibinfo  {journal} {Optics express}\
  }\textbf {\bibinfo {volume} {18}},\ \bibinfo {pages} {20143} (\bibinfo {year}
  {2010})}\BibitemShut {NoStop}%
\bibitem [{\citenamefont {Lin}\ and\ \citenamefont
  {Agrawal}(2006)}]{lin2006silicon}%
  \BibitemOpen
  \bibfield  {author} {\bibinfo {author} {\bibfnamefont {Q.}~\bibnamefont
  {Lin}}\ and\ \bibinfo {author} {\bibfnamefont {G.~P.}\ \bibnamefont
  {Agrawal}},\ }\href@noop {} {\bibfield  {journal} {\bibinfo  {journal}
  {Optics Letters}\ }\textbf {\bibinfo {volume} {31}},\ \bibinfo {pages} {3140}
  (\bibinfo {year} {2006})}\BibitemShut {NoStop}%
\bibitem [{\citenamefont {Lin}\ \emph {et~al.}(2007)\citenamefont {Lin},
  \citenamefont {Painter},\ and\ \citenamefont {Agrawal}}]{lin2007nonlinear}%
  \BibitemOpen
  \bibfield  {author} {\bibinfo {author} {\bibfnamefont {Q.}~\bibnamefont
  {Lin}}, \bibinfo {author} {\bibfnamefont {O.~J.}\ \bibnamefont {Painter}}, \
  and\ \bibinfo {author} {\bibfnamefont {G.~P.}\ \bibnamefont {Agrawal}},\
  }\href@noop {} {\bibfield  {journal} {\bibinfo  {journal} {Optics Express}\
  }\textbf {\bibinfo {volume} {15}},\ \bibinfo {pages} {16604} (\bibinfo {year}
  {2007})}\BibitemShut {NoStop}%
\bibitem [{\citenamefont {Osgood~Jr}\ \emph {et~al.}(2009)\citenamefont
  {Osgood~Jr}, \citenamefont {Panoiu}, \citenamefont {Dadap}, \citenamefont
  {Liu}, \citenamefont {Chen}, \citenamefont {Hsieh}, \citenamefont {Dulkeith},
  \citenamefont {Green}, \citenamefont {Vlasov} \emph
  {et~al.}}]{osgood2009engineering}%
  \BibitemOpen
  \bibfield  {author} {\bibinfo {author} {\bibfnamefont {R.}~\bibnamefont
  {Osgood~Jr}}, \bibinfo {author} {\bibfnamefont {N.}~\bibnamefont {Panoiu}},
  \bibinfo {author} {\bibfnamefont {J.}~\bibnamefont {Dadap}}, \bibinfo
  {author} {\bibfnamefont {X.}~\bibnamefont {Liu}}, \bibinfo {author}
  {\bibfnamefont {X.}~\bibnamefont {Chen}}, \bibinfo {author} {\bibfnamefont
  {I.-W.}\ \bibnamefont {Hsieh}}, \bibinfo {author} {\bibfnamefont
  {E.}~\bibnamefont {Dulkeith}}, \bibinfo {author} {\bibfnamefont
  {W.}~\bibnamefont {Green}}, \bibinfo {author} {\bibfnamefont
  {Y.}~\bibnamefont {Vlasov}},  \emph {et~al.},\ }\href@noop {} {\bibfield
  {journal} {\bibinfo  {journal} {Advances in Optics and Photonics}\ }\textbf
  {\bibinfo {volume} {1}},\ \bibinfo {pages} {162} (\bibinfo {year}
  {2009})}\BibitemShut {NoStop}%
\bibitem [{\citenamefont {Chen}\ \emph {et~al.}(2011)\citenamefont {Chen},
  \citenamefont {Levine}, \citenamefont {Fan},\ and\ \citenamefont
  {Migdall}}]{chen2011frequency}%
  \BibitemOpen
  \bibfield  {author} {\bibinfo {author} {\bibfnamefont {J.}~\bibnamefont
  {Chen}}, \bibinfo {author} {\bibfnamefont {Z.~H.}\ \bibnamefont {Levine}},
  \bibinfo {author} {\bibfnamefont {J.}~\bibnamefont {Fan}}, \ and\ \bibinfo
  {author} {\bibfnamefont {A.~L.}\ \bibnamefont {Migdall}},\ }\href@noop {}
  {\bibfield  {journal} {\bibinfo  {journal} {Optics Express}\ }\textbf
  {\bibinfo {volume} {19}},\ \bibinfo {pages} {1470} (\bibinfo {year}
  {2011})}\BibitemShut {NoStop}%
\bibitem [{\citenamefont {Sharping}\ \emph {et~al.}(2006)\citenamefont
  {Sharping}, \citenamefont {Lee}, \citenamefont {Foster}, \citenamefont
  {Turner}, \citenamefont {Schmidt}, \citenamefont {Lipson}, \citenamefont
  {Gaeta},\ and\ \citenamefont {Kumar}}]{sharping2006generation}%
  \BibitemOpen
  \bibfield  {author} {\bibinfo {author} {\bibfnamefont {J.~E.}\ \bibnamefont
  {Sharping}}, \bibinfo {author} {\bibfnamefont {K.~F.}\ \bibnamefont {Lee}},
  \bibinfo {author} {\bibfnamefont {M.~A.}\ \bibnamefont {Foster}}, \bibinfo
  {author} {\bibfnamefont {A.~C.}\ \bibnamefont {Turner}}, \bibinfo {author}
  {\bibfnamefont {B.~S.}\ \bibnamefont {Schmidt}}, \bibinfo {author}
  {\bibfnamefont {M.}~\bibnamefont {Lipson}}, \bibinfo {author} {\bibfnamefont
  {A.~L.}\ \bibnamefont {Gaeta}}, \ and\ \bibinfo {author} {\bibfnamefont
  {P.}~\bibnamefont {Kumar}},\ }\href@noop {} {\bibfield  {journal} {\bibinfo
  {journal} {Optics Express}\ }\textbf {\bibinfo {volume} {14}},\ \bibinfo
  {pages} {12388} (\bibinfo {year} {2006})}\BibitemShut {NoStop}%
\bibitem [{\citenamefont {Takesue}\ \emph
  {et~al.}(2008{\natexlab{a}})\citenamefont {Takesue}, \citenamefont {Fukuda},
  \citenamefont {Tsuchizawa}, \citenamefont {Watanabe}, \citenamefont {Yamada},
  \citenamefont {Tokura},\ and\ \citenamefont
  {Itabashi}}]{takesue2008entanglement}%
  \BibitemOpen
  \bibfield  {author} {\bibinfo {author} {\bibfnamefont {H.}~\bibnamefont
  {Takesue}}, \bibinfo {author} {\bibfnamefont {H.}~\bibnamefont {Fukuda}},
  \bibinfo {author} {\bibfnamefont {T.}~\bibnamefont {Tsuchizawa}}, \bibinfo
  {author} {\bibfnamefont {T.}~\bibnamefont {Watanabe}}, \bibinfo {author}
  {\bibfnamefont {K.}~\bibnamefont {Yamada}}, \bibinfo {author} {\bibfnamefont
  {Y.}~\bibnamefont {Tokura}}, \ and\ \bibinfo {author} {\bibfnamefont {S.-I.}\
  \bibnamefont {Itabashi}},\ }in\ \href@noop {} {\emph {\bibinfo {booktitle}
  {Group IV Photonics, 2008 5th IEEE International Conference on}}}\ (\bibinfo
  {organization} {IEEE},\ \bibinfo {year} {2008})\ pp.\ \bibinfo {pages}
  {404--406}\BibitemShut {NoStop}%
\bibitem [{\citenamefont {Takesue}\ \emph
  {et~al.}(2008{\natexlab{b}})\citenamefont {Takesue}, \citenamefont {Fukuda},
  \citenamefont {Tsuchizawa}, \citenamefont {Watanabe}, \citenamefont {Yamada},
  \citenamefont {Tokura},\ and\ \citenamefont
  {Itabashi}}]{takesue2008generation}%
  \BibitemOpen
  \bibfield  {author} {\bibinfo {author} {\bibfnamefont {H.}~\bibnamefont
  {Takesue}}, \bibinfo {author} {\bibfnamefont {H.}~\bibnamefont {Fukuda}},
  \bibinfo {author} {\bibfnamefont {T.}~\bibnamefont {Tsuchizawa}}, \bibinfo
  {author} {\bibfnamefont {T.}~\bibnamefont {Watanabe}}, \bibinfo {author}
  {\bibfnamefont {K.}~\bibnamefont {Yamada}}, \bibinfo {author} {\bibfnamefont
  {Y.}~\bibnamefont {Tokura}}, \ and\ \bibinfo {author} {\bibfnamefont {S.-i.}\
  \bibnamefont {Itabashi}},\ }\href@noop {} {\bibfield  {journal} {\bibinfo
  {journal} {Optics Express}\ }\textbf {\bibinfo {volume} {16}},\ \bibinfo
  {pages} {5721} (\bibinfo {year} {2008}{\natexlab{b}})}\BibitemShut {NoStop}%
\bibitem [{\citenamefont {Harada}\ \emph {et~al.}(2008)\citenamefont {Harada},
  \citenamefont {Takesue}, \citenamefont {Fukuda}, \citenamefont {Tsuchizawa},
  \citenamefont {Watanabe}, \citenamefont {Yamada}, \citenamefont {Tokura},\
  and\ \citenamefont {Itabashi}}]{harada2008generation}%
  \BibitemOpen
  \bibfield  {author} {\bibinfo {author} {\bibfnamefont {K.-i.}\ \bibnamefont
  {Harada}}, \bibinfo {author} {\bibfnamefont {H.}~\bibnamefont {Takesue}},
  \bibinfo {author} {\bibfnamefont {H.}~\bibnamefont {Fukuda}}, \bibinfo
  {author} {\bibfnamefont {T.}~\bibnamefont {Tsuchizawa}}, \bibinfo {author}
  {\bibfnamefont {T.}~\bibnamefont {Watanabe}}, \bibinfo {author}
  {\bibfnamefont {K.}~\bibnamefont {Yamada}}, \bibinfo {author} {\bibfnamefont
  {Y.}~\bibnamefont {Tokura}}, \ and\ \bibinfo {author} {\bibfnamefont {S.-i.}\
  \bibnamefont {Itabashi}},\ }\href@noop {} {\bibfield  {journal} {\bibinfo
  {journal} {Optics Express}\ }\textbf {\bibinfo {volume} {16}},\ \bibinfo
  {pages} {20368} (\bibinfo {year} {2008})}\BibitemShut {NoStop}%
\bibitem [{\citenamefont {Harada}\ \emph {et~al.}(2010)\citenamefont {Harada},
  \citenamefont {Takesue}, \citenamefont {Fukuda}, \citenamefont {Tsuchizawa},
  \citenamefont {Watanabe}, \citenamefont {Yamada}, \citenamefont {Tokura},\
  and\ \citenamefont {Itabashi}}]{harada2010frequency}%
  \BibitemOpen
  \bibfield  {author} {\bibinfo {author} {\bibfnamefont {K.-i.}\ \bibnamefont
  {Harada}}, \bibinfo {author} {\bibfnamefont {H.}~\bibnamefont {Takesue}},
  \bibinfo {author} {\bibfnamefont {H.}~\bibnamefont {Fukuda}}, \bibinfo
  {author} {\bibfnamefont {T.}~\bibnamefont {Tsuchizawa}}, \bibinfo {author}
  {\bibfnamefont {T.}~\bibnamefont {Watanabe}}, \bibinfo {author}
  {\bibfnamefont {K.}~\bibnamefont {Yamada}}, \bibinfo {author} {\bibfnamefont
  {Y.}~\bibnamefont {Tokura}}, \ and\ \bibinfo {author} {\bibfnamefont {S.-i.}\
  \bibnamefont {Itabashi}},\ }\href@noop {} {\bibfield  {journal} {\bibinfo
  {journal} {Selected Topics in Quantum Electronics, IEEE Journal of}\ }\textbf
  {\bibinfo {volume} {16}},\ \bibinfo {pages} {325} (\bibinfo {year}
  {2010})}\BibitemShut {NoStop}%
\bibitem [{\citenamefont {Coelho}\ \emph {et~al.}(2009)\citenamefont {Coelho},
  \citenamefont {Barbosa}, \citenamefont {Cassemiro}, \citenamefont {Villar},
  \citenamefont {Martinelli},\ and\ \citenamefont
  {Nussenzveig}}]{coelho2009three}%
  \BibitemOpen
  \bibfield  {author} {\bibinfo {author} {\bibfnamefont {A.}~\bibnamefont
  {Coelho}}, \bibinfo {author} {\bibfnamefont {F.}~\bibnamefont {Barbosa}},
  \bibinfo {author} {\bibfnamefont {K.}~\bibnamefont {Cassemiro}}, \bibinfo
  {author} {\bibfnamefont {A.}~\bibnamefont {Villar}}, \bibinfo {author}
  {\bibfnamefont {M.}~\bibnamefont {Martinelli}}, \ and\ \bibinfo {author}
  {\bibfnamefont {P.}~\bibnamefont {Nussenzveig}},\ }\href@noop {} {\bibfield
  {journal} {\bibinfo  {journal} {Science}\ }\textbf {\bibinfo {volume}
  {326}},\ \bibinfo {pages} {823} (\bibinfo {year} {2009})}\BibitemShut
  {NoStop}%
\bibitem [{\citenamefont {Matsko}\ \emph {et~al.}(2005)\citenamefont {Matsko},
  \citenamefont {Savchenkov}, \citenamefont {Strekalov}, \citenamefont
  {Ilchenko},\ and\ \citenamefont {Maleki}}]{matsko2005optical}%
  \BibitemOpen
  \bibfield  {author} {\bibinfo {author} {\bibfnamefont {A.~B.}\ \bibnamefont
  {Matsko}}, \bibinfo {author} {\bibfnamefont {A.~A.}\ \bibnamefont
  {Savchenkov}}, \bibinfo {author} {\bibfnamefont {D.}~\bibnamefont
  {Strekalov}}, \bibinfo {author} {\bibfnamefont {V.~S.}\ \bibnamefont
  {Ilchenko}}, \ and\ \bibinfo {author} {\bibfnamefont {L.}~\bibnamefont
  {Maleki}},\ }\href@noop {} {\bibfield  {journal} {\bibinfo  {journal}
  {Physical Review A}\ }\textbf {\bibinfo {volume} {71}},\ \bibinfo {pages}
  {033804} (\bibinfo {year} {2005})}\BibitemShut {NoStop}%
\bibitem [{\citenamefont {Driscoll}\ \emph {et~al.}(2009)\citenamefont
  {Driscoll}, \citenamefont {Liu}, \citenamefont {Yasseri}, \citenamefont
  {Hsieh}, \citenamefont {Dadap},\ and\ \citenamefont
  {Osgood}}]{driscoll2009large}%
  \BibitemOpen
  \bibfield  {author} {\bibinfo {author} {\bibfnamefont {J.~B.}\ \bibnamefont
  {Driscoll}}, \bibinfo {author} {\bibfnamefont {X.}~\bibnamefont {Liu}},
  \bibinfo {author} {\bibfnamefont {S.}~\bibnamefont {Yasseri}}, \bibinfo
  {author} {\bibfnamefont {I.}~\bibnamefont {Hsieh}}, \bibinfo {author}
  {\bibfnamefont {J.~I.}\ \bibnamefont {Dadap}}, \ and\ \bibinfo {author}
  {\bibfnamefont {R.~M.}\ \bibnamefont {Osgood}},\ }\href@noop {} {\bibfield
  {journal} {\bibinfo  {journal} {Optics Express}\ }\textbf {\bibinfo {volume}
  {17}},\ \bibinfo {pages} {2797} (\bibinfo {year} {2009})}\BibitemShut
  {NoStop}%
\bibitem [{\citenamefont {Fallahkhair}\ \emph {et~al.}(2008)\citenamefont
  {Fallahkhair}, \citenamefont {Li},\ and\ \citenamefont
  {Murphy}}]{fallahkhair2008vector}%
  \BibitemOpen
  \bibfield  {author} {\bibinfo {author} {\bibfnamefont {A.~B.}\ \bibnamefont
  {Fallahkhair}}, \bibinfo {author} {\bibfnamefont {K.~S.}\ \bibnamefont {Li}},
  \ and\ \bibinfo {author} {\bibfnamefont {T.~E.}\ \bibnamefont {Murphy}},\
  }\href@noop {} {\bibfield  {journal} {\bibinfo  {journal} {Journal of
  Lightwave Technology}\ }\textbf {\bibinfo {volume} {26}},\ \bibinfo {pages}
  {1423} (\bibinfo {year} {2008})}\BibitemShut {NoStop}%
\bibitem [{\citenamefont {Gardiner}(1985)}]{gardiner1985stochastic}%
  \BibitemOpen
  \bibfield  {author} {\bibinfo {author} {\bibfnamefont {C.}~\bibnamefont
  {Gardiner}},\ }\href@noop {} {\emph {\bibinfo {title} {Stochastic Methods}}}\
  (\bibinfo  {publisher} {Springer-Verlag, Berlin--Heidelberg--New
  York--Tokyo},\ \bibinfo {year} {1985})\BibitemShut {NoStop}%
\bibitem [{\citenamefont {Gardiner}\ and\ \citenamefont
  {Collett}(1985)}]{gardiner1985input}%
  \BibitemOpen
  \bibfield  {author} {\bibinfo {author} {\bibfnamefont {C.~W.}\ \bibnamefont
  {Gardiner}}\ and\ \bibinfo {author} {\bibfnamefont {M.~J.}\ \bibnamefont
  {Collett}},\ }\href@noop {} {\bibfield  {journal} {\bibinfo  {journal}
  {Physical Review A}\ }\textbf {\bibinfo {volume} {31}},\ \bibinfo {pages}
  {3761} (\bibinfo {year} {1985})}\BibitemShut {NoStop}%
\bibitem [{\citenamefont {van Loock}\ and\ \citenamefont
  {Furusawa}(2003)}]{loock2003detecting}%
  \BibitemOpen
  \bibfield  {author} {\bibinfo {author} {\bibfnamefont {P.}~\bibnamefont {van
  Loock}}\ and\ \bibinfo {author} {\bibfnamefont {A.}~\bibnamefont
  {Furusawa}},\ }\href {\doibase 10.1103/PhysRevA.67.052315} {\bibfield
  {journal} {\bibinfo  {journal} {Physical Review A}\ }\textbf {\bibinfo
  {volume} {67}},\ \bibinfo {pages} {052315} (\bibinfo {year}
  {2003})}\BibitemShut {NoStop}%
\end{thebibliography}%

\end{document}